\documentstyle[12pt,epsfig]{article}
\newcommand{\nc}{\newcommand}

\def\sb{s_\beta}
\def\cb{c_\beta}
\def\lam{\lambda}
\def\gam{\gamma}
\nc{\fdiag}{0}
\nc{\bg}{B. Grzadkowski}
\nc{\BG}{Bohdan Grzadkowski}
\nc{\lsp}{\;\;\;\;\;\;\;\;}
\nc{\beq}{\begin{equation}}   \nc{\eeq}{\end{equation}}
\nc{\bea}{\begin{eqnarray}}   \nc{\eea}{\end{eqnarray}}
\nc{\baa}{\begin{array}}      \nc{\eaa}{\end{array}}
\nc{\bit}{\begin{itemize}}    \nc{\eit}{\end{itemize}}
\nc{\ben}{\begin{enumerate}}  \nc{\een}{\end{enumerate}}
\nc{\bce}{\begin{center}}     \nc{\ece}{\end{center}}
\nc{\non}{\nonumber}
\nc{\lumun}{\;{\hbox {fb}^{-1}}{\hbox {yr}^{-1}}}
\nc{\hc}{\hbox {h.c.}}
\nc{\re}{\hbox {Re}}
\nc{\im}{\hbox {Im}}
\nc{\etal}{\hbox{et al.}}
\nc{\pbarn}{\;\hbox {pb}}
\nc{\prd}[3]{{ Phys.\ Rev.}\ {{\bf D{#1}}, #3 (#2)}}
\nc{\prl}[3]{{ Phys.\ Rev.\ Lett.}\ {{\bf {#1}}, #3 (#2)}}
\nc{\plb}[3]{{ Phys.\ Lett.}\ {{\bf B{#1}}, #3 (#2)}}
\nc{\npb}[3]{{ Nucl.\ Phys.}\ {{\bf B{#1}}, #3 (#2)}}
\nc{\ptp}[3]{{ Prog.\ Theor.\ Phys.}\ {{\bf {#1}}, #3 (#2)}}
\nc{\zfp}[3]{{ Z.\ Phys.}\ {{\bf C{#1}}, #3 (#2)}}
\nc{\mpla}[3]{{ Mod.\ Phys.\ Lett.}\ {{\bf A{#1}}, #3 (#2)}}
\nc{\rmp}[3]{{ Rev.\ Mod.\ Phys.}\ {{\bf {#1}}, #3 (#2)}}
\nc{\ijmpa}[3]{{ Int.\ J.\ of\ Mod.\ Phys.}\
               {{\bf A{#1}}, #3 (#2)}}
\nc{\app}[3]{{ Acta\ Phys.\ Polon.}\ {{\bf B{#1}}, #3 (#2)}}
\nc{\epj}[3]{{ Eur. Phys. J.}\ {{\bf C{#1}}, #3 (#2)}}
\nc{\prep}[3]{{ Phys.\ Rep.}\ {{\bf {#1}}, #3 (#2)}}
\nc{\ra} {\rightarrow}
\nc{\cw}{\cos\theta_W}        
\nc{\sw}{\sin\theta_W}
\nc{\ttbar}{t\bar{t}}
\nc{\bbbar}{b\bar{b}}
\nc{\tanb} {\tan \beta}
\nc{\cotb} {\cot\beta}
\nc{\twbdec} {t\rightarrow W^+ b}
\nc{\tbwbdec} {\bar{t} \rightarrow W^- \bar{b}}
\nc{\hprod} {e^+e^- \ra Z^\ast \ra H Z}
\nc{\epem} {e^+e^-}
\nc{\wpwm} {W^+W^-}
\nc{\tbar} {\bar{t}}
\nc{\bbar} {\bar{b}}

\nc{\wpp} {W^+}
\nc{\mt}{m_t}
\nc{\mts}{m_t^2}
\nc{\mw} {m_W}
\nc{\mws} {m_W^2}
\nc{\mz} {m_Z}
\nc{\mzs} {m_Z^2}
\nc{\mhs} {m_H^2}
\nc{\ma} {m_A}
\nc{\mas} {m_A^2}
\nc{\hdec}{H \ra t\bar{t}}
\nc{\ttbardec}{\ttbar \ra W^+W^-\bbbar}
\nc{\po}{\Phi_1}
\nc{\pod}{\Phi_1^\dagger}
\nc{\pht}{\Phi_2}
\nc{\phtd}{\Phi_2^\dagger}
\nc{\phtt}{{\tilde{\Phi}}_2}
\nc{\popo}{\po^\dagger\po}
\nc{\phtpt}{\pht^\dagger\pht}
\nc{\popt}{\po^\dagger\pht}
\nc{\phtpo}{\pht^\dagger\po}
\nc{\sq}{\sqrt{2}}
\nc{\nsd} {N_{SD}}
\nc{\ntt} {N_{tt}}
\nc{\vs}{\vspace{2mm}}

\nc{\sty}{\hat{S}^t_1} \nc{\pty}{\hat{P}^t_1}
\nc{\sts}{(\sty)^2}      \nc{\pts}{(\pty)^2}
\nc{\yts}{\sts+\pts}
\nc{\sby}{\hat{S}^b_1} \nc{\pby}{\hat{P}^b_1}
\nc{\sbs}{(\sby)^2}      \nc{\pbs}{(\pby)^2}
\nc{\ybs}{\sbs+\pbs}

\nc{\styi}{\hat{S}^t_i} \nc{\ptyi}{\hat{P}^t_i}
\nc{\stsi}{(\styi)^2}      \nc{\ptsi}{(\ptyi)^2}
\nc{\ytsi}{\stsi+\ptsi}
\nc{\sbyi}{\hat{S}^b_i} \nc{\pbyi}{\hat{P}^b_i}
\nc{\sbsi}{(\sbyi)^2}      \nc{\pbsi}{(\pbyi)^2}
\nc{\ybsi}{\sbsi+\pbsi}

\def\sa{s_\alpha}
\def\ca{c_\alpha}
\def\sb{s_\beta}
\def\cb{c_\beta}
\def\dchisq{\Delta\chi^2}
\def\dchisqmin{\dchisq_{\rm min}}

\def\vev#1{\langle #1 \rangle}

\def\sb{s_\beta}
\def\cb{c_\beta}
\def\rts{\sqrt s}

\def\hsm{h_{\rm SM}}
\def\mhsm{m_{\hsm}}

\def\h{h}
\def\mh{m_{\h}}
\def\H{H}
\def\mH{m_{\H}}

\def\hl{h^0}
\def\hh{H^0}
\def\ha{A^0}
\def\hp{H^+}
\def\hm{H^-}
\def\hpm{H^{\pm}}
\def\mhl{m_{\hl}}
\def\mhh{m_{\hh}}
\def\mha{m_{\ha}}

\def\mhpm{m_{\hpm}}
\def\lsim{\mathrel{\raise.3ex\hbox{$<$\kern-.75em\lower1ex\hbox{$\sim$}}}}
\def\gsim{\mathrel{\raise.3ex\hbox{$>$\kern-.75em\lower1ex\hbox{$\sim$}}}}
\def\anti{\overline}

\def\fbi{~{\rm fb}^{-1}}

\def\gev{\,{\rm GeV}}
\def\tev{\,{\rm TeV}}

\begin{document}
%
\font\fortssbx=cmssbx10 scaled \magstep2
\medskip
\begin{flushright}
$\vcenter{
\hbox{\bf UCD-2000-15} 
\hbox{\bf IFT/00-18}
\hbox{\bf hep-ph/0009271}
\hbox{September, 2000}
}$
\end{flushright}
\vspace*{.7cm}
\begin{center}
{\large{\bf Do precision electroweak constraints guarantee {\boldmath
$\epem$} collider discovery of at least one Higgs boson of a
two-Higgs-doublet model?}}\\ \rm
\vspace*{1cm}
\renewcommand{\thefootnote}{\alph{footnote})}
{\bf Piotr Chankowski$^1$}
\footnote{E-mail:{\tt chank@fuw.edu.pl}}
{\bf Thomas Farris$^2$}
\footnote{E-mail:{\tt farris@physics.ucdavis.edu}}
{\bf Bohdan Grzadkowski$^1$}
\footnote{E-mail:{\tt bohdang@fuw.edu.pl}} 
{\bf John F. Gunion$^2$}
\footnote{E-mail:{\tt jfgucd@higgs.ucdavis.edu}} 
{\bf Jan Kalinowski$^1$}
\footnote{E-mail:{\tt kalino@fuw.edu.pl}}
{\bf Maria Krawczyk$^1$}
\footnote{E-mail: {\tt krawczyk@fuw.edu.pl}}

\vspace*{.4cm}
{$^1$ \em Instytut Fizyki Teoretycznej UW, Hoza 69, 
Warsaw, Poland}\\
{$^2$ \em Davis Institute for High Energy Physics, 
UC Davis, CA, USA }\\

\vspace*{.7cm}

{\bf Abstract}

\end{center}
\vspace{5mm}    We consider a CP-conserving
two-Higgs-doublet type II model with a light scalar or 
pseudoscalar neutral Higgs boson
($\h=\hl$ or $\h=\ha$) that has no $ZZ/WW$ coupling and,
thus, cannot be detected in $\epem\to Z\h$ (Higgs-strahlung) 
or $\nu\anti\nu \h$ (via $WW$ fusion).
Despite sum rules which ensure that the light $\h$ must
have significant $t\anti t$ or $b\anti b$ coupling,
for a wedge of moderate $\tanb$, that becomes increasingly large
as $\mh$ increases, the $\h$
can also escape discovery in both $b\anti b \h$
and $t\anti t \h$ production at a $\rts=500-800\gev$ $\epem$ collider 
(for expected luminosities). If the other Higgs
bosons happen to be too heavy to be produced,
then no Higgs boson would be detected. We demonstrate that,
despite such high masses for the other Higgs bosons, only the low-$\tanb$
portion of the no-discovery wedges in $[\mh,\tanb]$ parameter
space can be excluded due to failure
to fit precision electroweak observables.  
In the $\tanb\gsim 1$  regions of the no-discovery wedges, 
we find that the 2HDM fit
to precision electroweak observables has small $\Delta\chi^2$
relative to the best minimal one-doublet SM fit.

\vfill
\clearpage

\renewcommand{\thefootnote}{\sharp\arabic{footnote}}
\setcounter{footnote}{0}

{\bf 1. Introduction.} Spontaneous gauge symmetry breaking in the
Standard Model (SM) is realized by introducing a single Higgs doublet
field, leading to a single CP-even Higgs
boson, $\hsm$. In this 1HDM, Higgs discovery at an $\epem$
collider is always possible via the Higgs-strahlung process, $\epem\to Z\hsm$
\cite{lcrep} provided only that $\rts>\mz+\mhsm$.  However, 
even the simplest CP-conserving two-Higgs-doublet
model (2HDM) extension of the SM exhibits a rich Higgs sector structure
that makes Higgs discovery more challenging.
The CP-conserving 2HDM predicts \cite{hhg}  the
existence of two neutral CP-even Higgs bosons ($\hl$ and $\hh$, with
$\mhl\leq\mhh$ by convention), one neutral CP-odd Higgs ($\ha$) 
and a charged Higgs pair ($\hpm$).  
We consider the type-II 2HDM,
wherein $\phi_1^0$ couples to down-type quarks and leptons
and $\phi_2^0$ to up-type quarks. We employ the conventions
$\vev{\phi_{1,2}^0}=v_{1,2}>0$, $\tanb=v_2/v_1$.
Tree-level Higgs coupling strengths relative
to SM-like couplings are then ($\sa\equiv \sin\alpha,\ldots$):
[$t\anti t$] $\hh,\hl,\ha\to \sa/\sb,\ca/\sb, \cb/\sb$;
[$b\anti b$] $\hh,\hl,\ha\to \ca/\cb,-\sa/\cb,\sb/\cb$;
[$WW,ZZ$] $\hh,\hl,\ha\to \cos(\alpha-\beta),\sin(\beta-\alpha),0$,
where the $\ha f\anti f$ couplings have an extra $i\gamma_5$
and the range $-\pi<\alpha\leq 0$  
covers the full physical parameter space in the $\mhl<\mhh$ 
convention.~\footnote{We employ the conventional definition of
$\alpha$ according to which 
$\hh=\cos\alpha\sqrt 2 \Re \phi_1^0+\sin\alpha\sqrt2 \Re \phi_2^0$ and 
$\hl=-\sin\alpha\sqrt 2\Re\phi_1^0+\cos\alpha \sqrt 2 \Re\phi_2^0 $.
The $0<\alpha\leq\pi$ range is equivalent to $-\pi<\alpha\leq 0$
under $\alpha\to \alpha+\pi$ which is equivalent to 
simultaneously redefining
the overall signs of the $\hl$ and $\hh$ states; this has no physical
consequences.}

In this note, we wish to address the extent to
which a light $\h$ with zero $ZZ/WW$ coupling 
(and, therefore, undetectable in the $Z\h$ and $\nu\anti\nu \h$ final states)
is guaranteed to be discovered in the $t\anti t\h$ and/or $b\anti b\h$
final states at an $\epem$ LC collider even if all other Higgs bosons 
are too heavy to be produced {\it and} if the model
is required to be consistent with precision electroweak constraints.
We consider the cases of: a) $\h=\ha$ (the tree-level $\ha ZZ$ and $\ha WW$
couplings are automatically zero); and b) $\h=\hl$ with the scalar sector
mixing angle chosen so that $\sin(\beta-\alpha)=0$ 
(to zero the $\hl ZZ$ and $\hl WW$ couplings).\footnote{We have checked that
the non-zero $ZZ/WW$ couplings induced at one-loop are too small
to produce a detectable $Z\h$ or $\nu\anti\nu \h$ signal. E.g., relative to the
$Z\h$ and $\nu\anti\nu\h$ cross sections for a SM-like $\h$, 
the suppressions for $\h=\ha$ are of order ${\alpha_W^2}{\mt^4
\over s\mz^2}\cot^2\beta$ and $\alpha_W^2\cot^2\beta$, respectively.}
In earlier work \cite{ggk}, we found that
there are dangerous wedge-shaped regions
in the $\mh$-$\tanb$ plane, characterized by a range of 
moderate $\tanb$ values that increases in size as $\mh$ increases, for which
the $\h$ cannot be seen at a future LC, even though it is guaranteed that
the $t\anti t \h$ and $b\anti b\h$ couplings-squared 
cannot be simultaneously suppressed when the $ZZ\h$ coupling is zero.
Even for a very high integrated 
luminosity, $L=2500\fbi$, such as might be achievable after several years
of running, at both $\rts=500\gev$ and $800\gev$ and $\mh$ 
well below $\rts-2m_t$, the
$t\anti t \h$ and $b\anti b\h$ production rates are simply
not adequate for detecting the $\h$.  However,
since the other Higgs bosons are assumed
to be quite heavy to avoid production, implying no {\it light}
Higgs with substantial $ZZ/WW$ couplings, it would seem that the
fit to precision electroweak constraints is likely to be poor
in the no-discovery wedges.
But, in \cite{maria} it was shown that a good global fit to EW
data is possible even for very light $\hl$ or $\ha$.
In this paper, we compute the best precision electroweak fit $\chi^2$'s
that can be achieved in the no-discovery wedges. Generally speaking,
for LC $\rts=500\gev$ ($800\gev$) we find that the $\tanb\gsim 2$ portions of
the 2HDM no-discovery wedges in $\mh$-$\tanb$ parameter space have
$\Delta\chi^2<1$ ($<1.5$) (relative to the best SM fit) and all of the
no-discovery wedges' portions with $\tanb\gsim 1$ have $\Delta\chi^2<2$.
The discrimination (following from current precision electroweak data)
between the minimal 1HDM SM and the no-discovery scenarios in the 2HDM
is thus rather weak for $\rts=500\gev$ and not all that much better
for $\rts=800\gev$. After presenting the results for the EW precision
fits and qualitative discussion we find an interestingly simple form
of the Higgs potential which ensures the smallest
$\Delta\chi^2$'s in the scenarios considered. 
Finally, we comment on extending the energy of the LC
and searches at the LHC.

\begin{figure}[h]
\begin{center}
{\epsfig{file=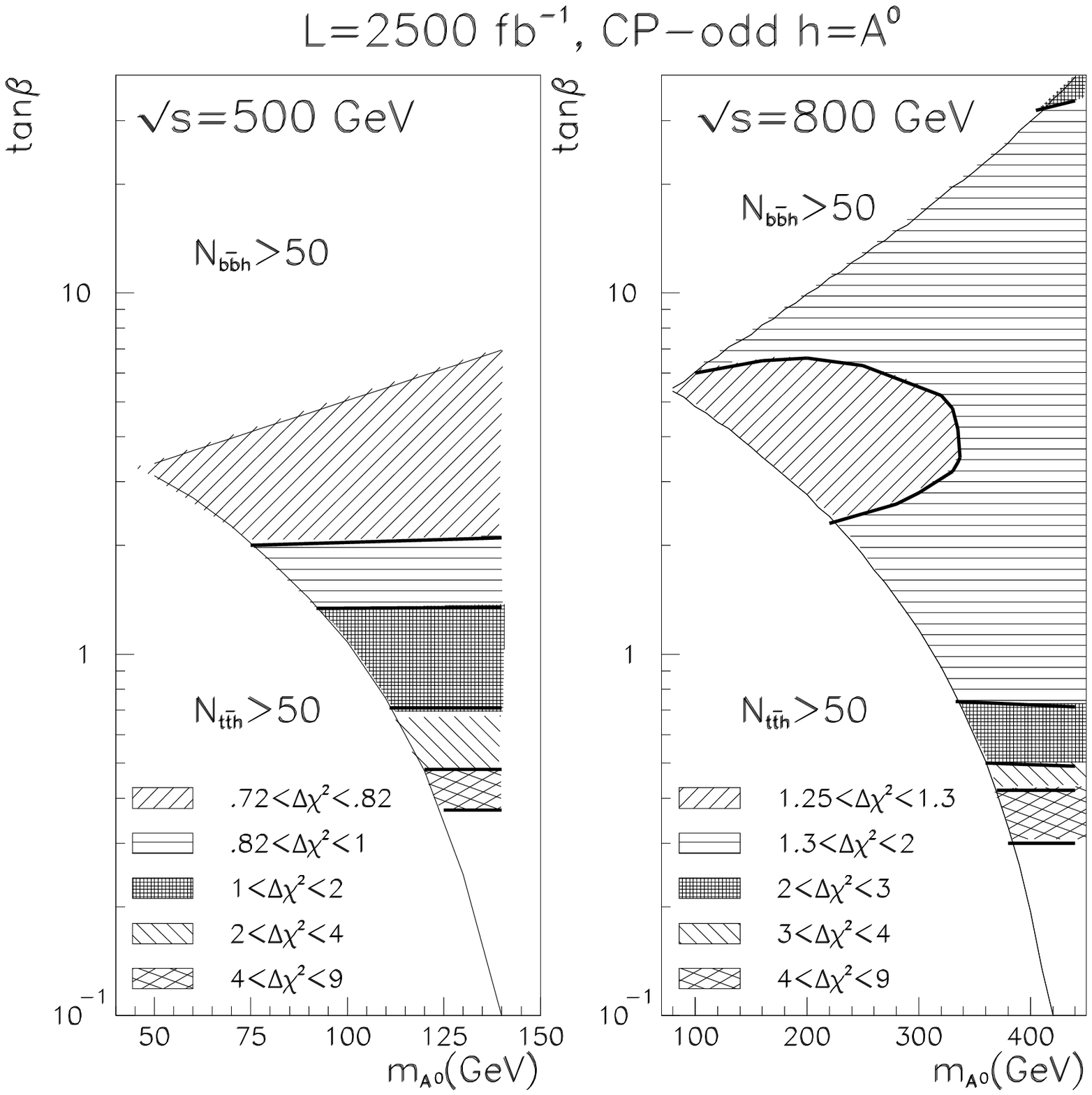,height=10cm,width=16cm}}
\caption{
For $\protect\rts=500\gev$  and $\protect\rts=800\gev$,
the solid lines show
as a function of $\mha$ the maximum and minimum $\tanb$
values between which $t\anti t \ha$, $b\anti b \ha$ 
final states will both have fewer than 50 events assuming
$L=2500\fbi$. The different types of bars indicate
the best $\chi^2$ values obtained for fits to precision electroweak
data after scanning: over the masses of the remaining Higgs bosons 
subject to the constraint they are 
too heavy to be directly produced; and over the mixing angle
in the CP-even sector.}
\label{wedgeha}
\end{center}
\end{figure}

\begin{figure}[h]
\begin{center}
{\epsfig{file=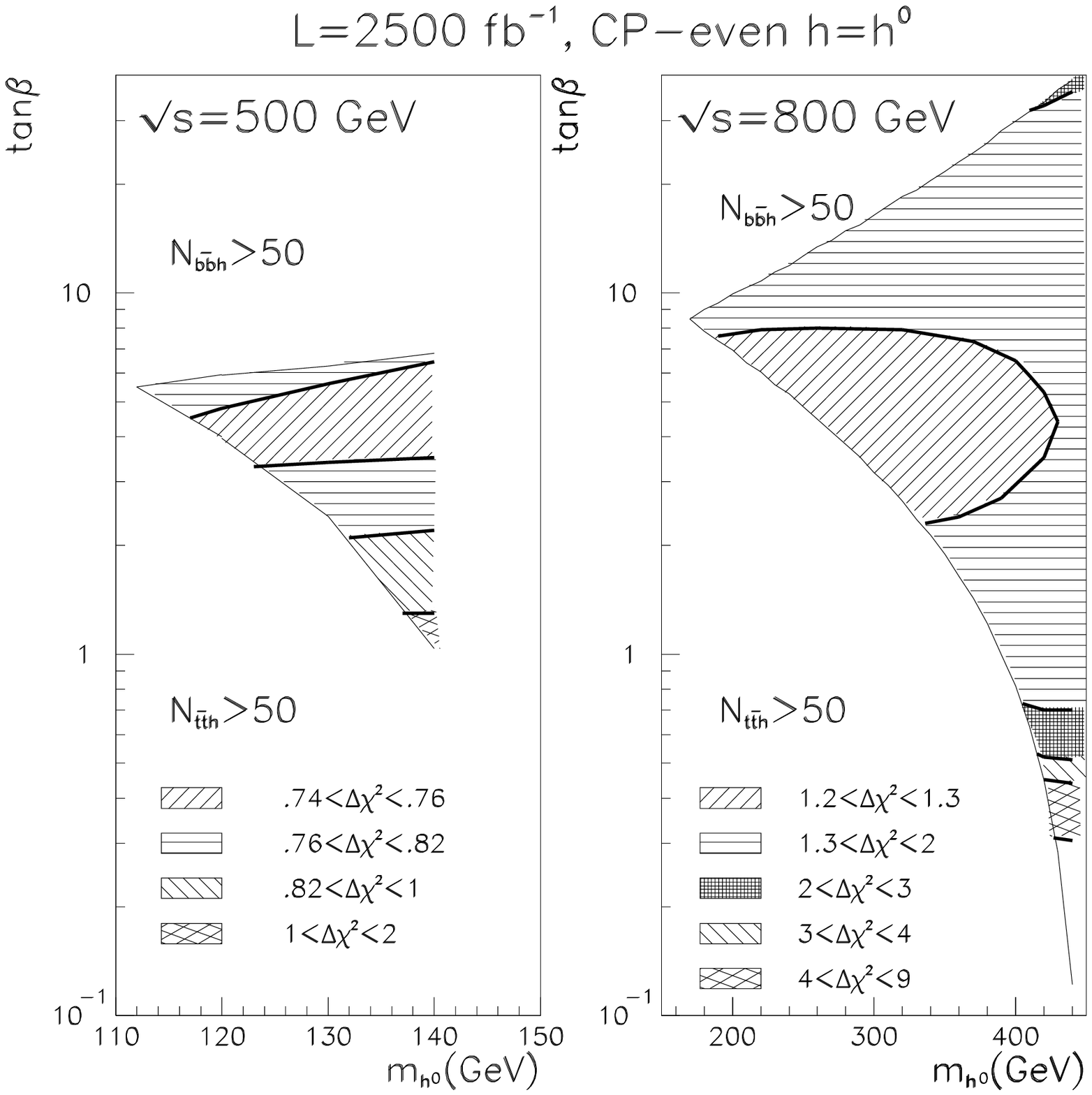,height=10cm,width=16cm}}
\caption{The same as for Fig.~\protect\ref{wedgeha}, except for $\h=\hl$.
The CP-even sector mixing angle
is fixed by the requirement $\sin(\beta-\alpha)=0$.}
\label{wedgehl}
\end{center}
\end{figure}

{\bf 2. No-discovery wedges.} We begin with a reminder and discussion
concerning the wedges of $\mh$-$\tanb$ parameter
space for which the $\h$ cannot be discovered. Our criterion
is that there should be fewer than 50 events in
 $t\anti t\h$ and in $b\anti b\h$
production assuming $L=2500\fbi$ and $\rts=500$ or $800\gev$.
(In practice, larger event numbers might turn out to be needed
for discovery; 50 events before cuts is probably a minimum.)
The wedges are shown by the solid black lines in Fig.~\ref{wedgeha}
for $\h=\ha$ and in Fig.~\ref{wedgehl}
for $\h=\hl$. These figures are confined
to $\mh$ values such that $t\anti t \h$ production is kinematically
allowed for the given $\rts$. The no-discovery zones extend beyond 
$\mh=\rts-2\mt$ with the difference that there is no lower-bound on $\tanb$.
However, low $\tanb$ values are highly 
disfavored by the precision electroweak constraints for all $\mh$. 
We shall discuss the case of $\mh\sim \rts$ later.
The $\h=\ha$ wedges are essentially the same as those found for the general
CP-violating 2HDM in \cite{ggk}. Even though the $\ha$ has exactly
the same magnitude (but not necessarily the same sign)
for its $\anti f i\gamma_5 f$ couplings 
as does the $\hl$ for its $\anti f 1 f$ couplings [for $\sin(\beta-\alpha)=0$],
the low (high) $\tanb$ limits of the $\h=\hl$ wedges 
are substantially above (close to) those for $\h=\ha$. This is
because the $f\anti f \hl$ cross section differs from the $f\anti f\ha$
cross section by a term proportional to $m_f^2$; the resulting difference
is substantial for $f=t$ (which determines
the low $\tanb$ limit of the wedge).

\begin{table}[htb]
\caption{Observables considered (TEV stands for Tevatron data) 
and typical pulls
for a 2HDM fit.  Pulls are defined as $({\cal O}_i-{\cal O}_i^{\rm min})
/\Delta{\cal O}_i$, where ${\cal O}_i$ is the measured value
of a given observable, ${\cal O}_i^{\rm min}$
is the value for the observable for the best fit choice of parameters,
and $\Delta{\cal O}_i$ is the full error (including systematic error) 
for that observable.
The pull results are for $\mt=174\gev$,
$\alpha_s=0.117$, $\mha=90\gev$, $\tanb=2.3$, $\mhl=490\gev$,
$\mhh=830\gev$ and $\mhpm=850\gev$, yielding $\dchisq=0.78$
relative to the best $\chi^2$ achieved in the SM-like limit of the 2HDM,
for which we also give the pulls
for the same $\mt$ and $\alpha_s$. These latter results are quite close
to those given in \cite{latestdata} with
the exception of $\mw^{\rm LEP}$ for which we have
used the Moriond result including LEP2 running \cite{mwlep2}. The SM-like 
2HDM pulls are essentially identical to those
of \cite{latestdata} if we use $\mw^{\rm LEP}$ as quoted there.
}
\label{pulls}
\renewcommand{\arraystretch}{1}
\medskip\centering
\begin{tabular}{|c|c|c|c|c|c|c|}
\hline
${\cal O}$ & $\mw^{\rm LEP}$ & $\mw^{\rm TEV}$ & $\sin^2\theta_W^{\rm TEV}$ 
& $\Gamma_{\rm tot}^Z$ &
$\sigma_{\rm had}^Z$ & ${\cal A}_e^{\rm LEP}$  \\
\hline
2HDM  & 0.157 & 0.880 & 1.32 & -0.972 & 1.61 & 0.338 \\
  SM  &  0.370 &  1.04 &  1.23 & -0.508 &  1.73  & 0.167 \\
\hline
\hline
${\cal O}$ & ${\cal A}_\tau^{\rm LEP}$ & $\sin^2\theta^*_{\rm LEP}$ 
& $\Gamma_{\rm had}^Z/\Gamma_{\rm lep}^Z$ & $A_{FB}^{l~{\rm LEP}}$ 
& $R_b^{\rm LEP}$ & $R_c^{\rm LEP}$   \\
\hline
2HDM & -0.927 & 0.522 &  1.42 & 0.944 & 0.733 & -0.744   \\
SM &  -1.12 &  0.632  & 1.13 & 0.742 &  0.668 & -0.743 \\
\hline
\hline
${\cal O}$ &  $A_{FB}^{b~{\rm LEP}}$ & $A_{FB}^{c~{\rm LEP}}$ & 
$A_{LR}^{b~{\rm SLD}}$ & $A_{LR}^{c~{\rm SLD}}$ & 
$\sin^2\theta_{\rm SLD}$ & \ \\
\hline
2HDM & -1.98 & -1.22 & -0.948 & -1.45 & -2.26 &\ \\
SM  &  -2.29 & -1.34 & -0.950 & -1.46 & -1.83 & \ \\
\hline 
\end{tabular} 
\end{table}

{\bf 3. Electroweak fits.} 
The question is the extent to which portions or all of these
wedges can be excluded because the model is inconsistent
with precision electroweak constraints when the masses of the
other Higgs bosons are taken to be sufficiently heavy
that they cannot be produced. For the heavy neutral (charged) Higgs bosons,  
the strongest constraint of this type is that $b\anti b H$ ($t b\hpm$) 
production be forbidden.~\footnote{If $b\anti b H$ is forbidden,
$\nu\anti\nu H$ will be extremely suppressed.}
Figs.~\ref{wedgeha} and \ref{wedgehl} show the best $\Delta\chi^2$ 
values obtained
for the precision electroweak fit at each $\tanb$ and $\mh$ ($=\mha$ or $\mhl$)
value in the $\rts=500$ and $800\gev$ no-discovery wedges after scanning
over all values of masses for the other (heavier)
neutral Higgs bosons above $\rts-10\gev$,
and over all $\hpm$ masses above $\rts-180\gev$.
In the case of $\h=\ha$, we also scan
over all mixing angles $\alpha$ in the CP-even sector.
The observables we consider are listed in Table~\ref{pulls}.
We employ the LEP/SLD/Tevatron experimental values from \cite{latestdata}
with the exception of $\mw^{\rm LEP}$ for which we 
take the LEP2 result \cite{mwlep2}. We
include statistical and systematic errors and correlations in computing 
the contribution to $\chi^2$ from each observable. 
We observe that rather small $\Delta\chi^2$ levels
are the rule once $\tanb\gsim 1$; $\Delta\chi^2<1$ ($<2$) is easily achieved
in the $\rts=500\gev$ ($800\gev$) wedges. 
For the no-discovery parameter regions, the 
best $\chi^2$ values for $\tanb> 0.5$ are always obtained 
in the $\h=\ha$ ($\hl$) cases, respectively, by taking 
the next lightest Higgs boson $\H$ 
to be $\H=\hl$ ($\hh$) with $\mhl$ ($\mhh$) as
light as consistent with forbidding $b\anti b \H$ production (i.e.,
$\mH=\rts-10\gev$).  
The large $\dchisq$'s found
for $\tanb$ values substantially below 1 derive primarily from
too large a result for $R_b$, although the deviation of
$\Gamma^Z_{\rm tot}$ from the observed value also becomes increasingly
large as $\tanb$ becomes small.

\begin{table}[h]
\caption{Lower and upper values of $\tanb$, using the notation  
$[\tanb_{\rm min},\tanb_{\rm max}]$, at which the given
$\dchisqmin$ value is crossed for the $\mh=\rts-10\gev$ cases.}
\label{boundaries}
\medskip
\begin{center}
\begin{tabular}{|c|c|c|c|c|c|}
\hline
$\dchisqmin$ & 1 & 2 & 3 & 4 & 9 \\
\hline
$\h=\ha,\rts=500$ & [1.8,14] & [0.63,56] & [0.49,75] & [0.44,89] & 
[0.30,$\;>110$] \\
$\h=\ha,\rts=800$ & no       & [0.75,47] & [0.46,85] & [0.39,107] & 
[0.27,$\;>110$] \\
$\h=\hl,\rts=500$ & no       & [0.92,51] & [0.73,73] & [0.63,86] & 
[0.45,$\;>110$] \\
$\h=\hl,\rts=800$ & no       & [1.4,33] & [0.68,78] & [0.55,102] & 
[0.35,$\;>110$] \\
\hline
\end{tabular}
\end{center}
\end{table}

Given that low $\tanb$ is disfavored by the precision electroweak fit,
it is appropriate to explore the $\tanb\gsim 0.5$ region out to values
of $\mh$ beyond the $\mh<\rts-2\mt$ no-discovery wedges considered 
originally in \cite{ggk}.  The figures make
clear that the no-discovery region for the $\h$ (whether $\ha$ or
$\hl$) reaches to increasingly higher $\tanb$ as $\mh$ increases. To
illustrate, we give additional results for $\mh=\rts-10\gev$, the
smallest $\mh$ for which $b\anti b$+Higgs is forbidden for all neutral Higgs
bosons. The boundaries in $\tanb$ for different
$\dchisqmin$ levels are given in Table~\ref{boundaries}.

To illustrate some additional points, we focus on the $\h=\ha$ case,
which also allows us to discuss issues related to choosing the scalar-sector
mixing angle $\alpha$.  Naively, it might be expected that the smallest
$\Delta\chi^2$ would be achieved when $\H=\hl$ has SM-like couplings to $ZZ,WW$
($\beta-\alpha\sim \pi/2$).  This is frequently the case, and, in such
cases, $\mhpm$ and $\mhh$ are typically much larger than $\mhl$,
while $\mhpm$ is closely correlated with and just slightly larger than $\mhh$.
However, another frequent $\dchisqmin$ configuration is $\alpha\sim 0$
(i.e. no mixing in the scalar sector).
For such cases, $\mhpm-\mhh$ is again not large and, in addition, $\mhh-\mhl$
is often small (sometimes $\mhh=\mhl$, in which case changing
$\alpha$ has no physical consequences). Further, 
$\dchisqmin$ cases with $\alpha\sim 0$
typically arise for $\tanb$ values for which $\sin^2(\beta-\alpha)$
is rather small ($<0.2$).  However, it is very common that the
smallest $\dchisq$ achieved with $\sin^2(\beta-\alpha)\sim 1$ is
only slightly larger than that found for $\alpha\sim 0$, and vice versa.
Indeed, for most $[\mha,\tanb]$ points in the no-discovery wedges,
$\dchisq$ values quite close to $\dchisqmin$ can be achieved for a rather
wide range of 2HDM parameters so long as $\mhpm-\mhh$ is positive but not large.

To illustrate in more detail, consider $\rts=500\gev$ and the
no-discovery point $[\mha=90\gev,\tanb=2.3]$, for which $\dchisqmin$ is
achieved for $\mhl=490\gev$ (i.e., as small as consistent
with absence of $b\anti b\hl$ production), $\mhh=830\gev$,
$\mhpm=850\gev$ and $\alpha=-0.1\pi$  (for which
$\beta-\alpha\sim \pi/2$, implying 
$\sin^2(\beta-\alpha)\sim 1$). 
The `pulls' of each of our 17 input observables, ${\cal O}_i$, 
for this $\dchisqmin$ parameter set are
listed in Table~\ref{pulls}. These values are typical. From the
table, we also see that
the 2HDM pattern of pulls is only somewhat different from the 
conventional 1HDM fit pulls obtained using the same input data,
with some ${\cal O}_i$ fit better in the 2HDM case than in the 1HDM
case and vice versa.

\begin{figure}[h]
\begin{center}
{\epsfig{file=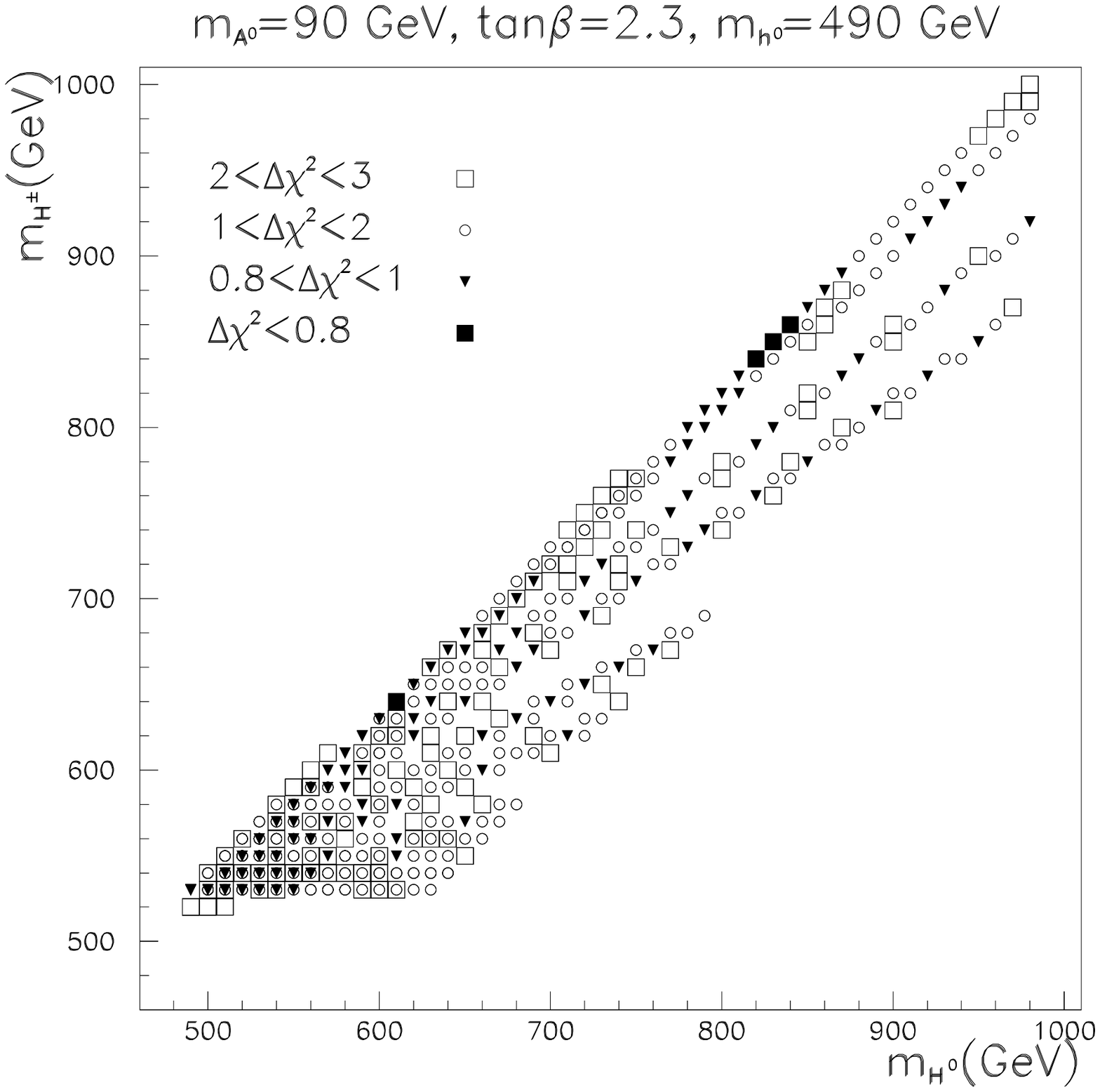,width=10cm}}
\end{center}
\caption{For the case of $\mha=90\gev$, $\tanb=2.3$ and $\mhl=490\gev$
(that yields the minimum $\dchisq$ for $\rts=500\gev$ when
requiring that $b\anti b\hl$ production is forbidden), 
we plot $\mhpm$  as a function of $\mhh$ for various
ranges of $\dchisq$. Scans in $\mhh$ and $\mhpm$
were done using 10 GeV steps, which leads to some incompleteness
in the points for each $\dchisq$ range. The scan in $\mhh$ was
limited to $\mhh<980\gev$. Multiple entries at the same $\mhh,\mhpm$
location correspond to different $\alpha$ values.}
\label{m3mc}
\end{figure}

\begin{figure}[p]
\begin{center}
{\epsfig{file=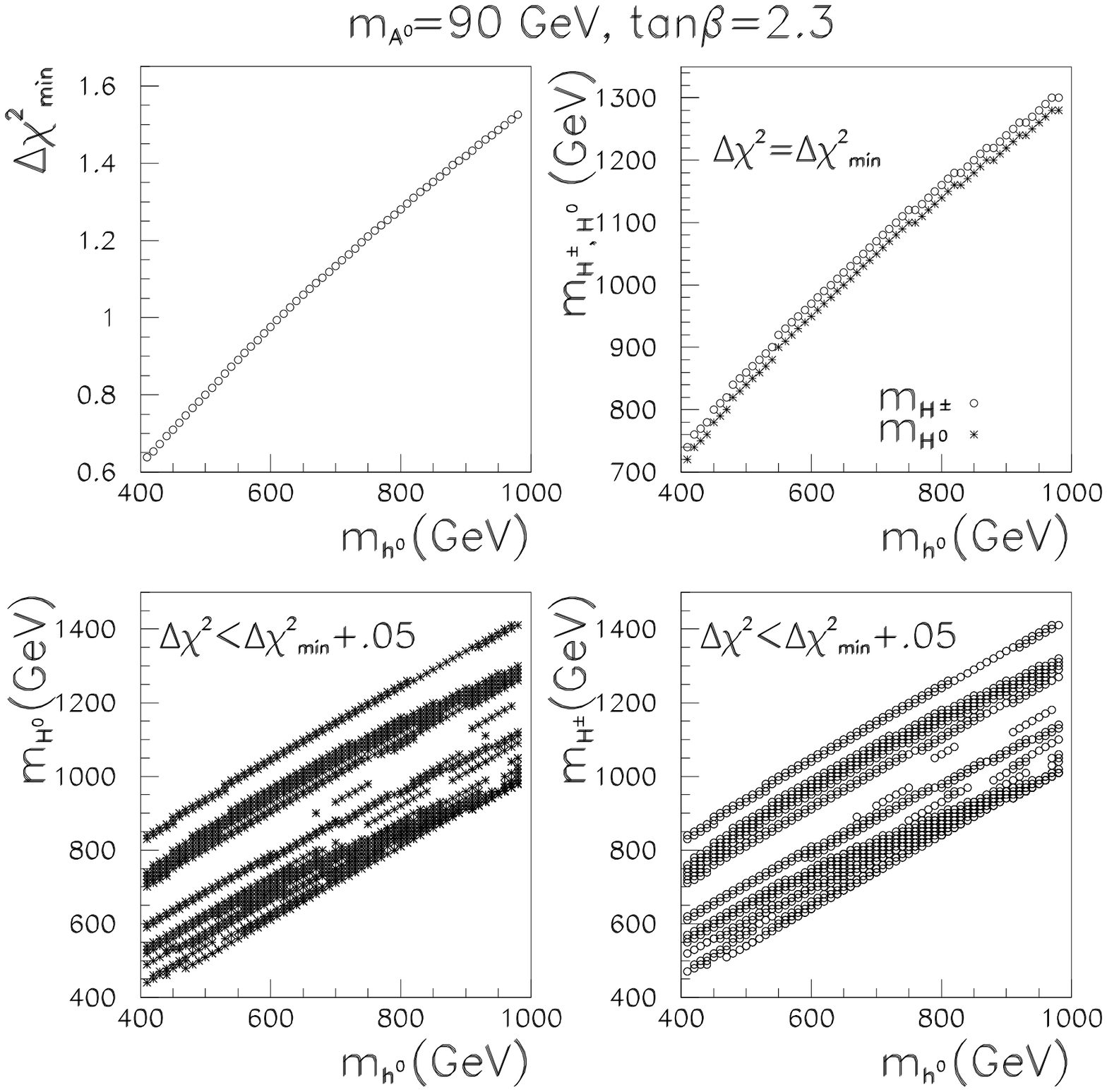,width=16cm}}
\end{center}
\smallskip
\caption{For the case of $\mha=90\gev$ and $\tanb=2.3$,
we plot as a function of $\mhl\in[410,980]\gev$: 
a) the minimum $\dchisq$ 
found after scanning over all values of $\mhh,\mhpm>\mhl$
and over all scalar sector mixing angles; b) the corresponding
values of $\mhh$ and $\mhpm$;
c) the values of $\mhh$ for which $\dchisq<\dchisqmin+0.05$ is achieved;
d) the closely correlated values of $\mhpm$ for which  
$\dchisq<\dchisqmin+0.05$ is achieved. For this case,
$\dchisqmin$ is always achieved for $\alpha=-0.1\pi$ 
(our scan is in units of $0.1\pi$), which roughly corresponds
to $\beta-\alpha=\pi/2$, i.e. maximal $\hl$ coupling to $ZZ$.
}
\label{chi2min}
\end{figure}

For $\mhl=490\gev$, a large range of values for $\mhh$ and $\mhpm$ is 
possible without increasing $\dchisq$ significantly, provided
$\mhpm$ is closely correlated with $\mhh$.  
This is illustrated in Fig.~\ref{m3mc}  which shows  
the values for $\mhpm$ as a function of $\mhh$
for $\Delta\chi^2<0.8$ and in the ranges [0.8,1], [1,2], and [2,3].
%
For the smallest $\dchisq$ values, $\mhpm-\mhh= 20\gev$.

Not surprisingly, there is an increase of minimum $\dchisq$
as $\mhl$ increases (for given $\mha$ and $\tanb$). Also,
the best choices of $\mhh$  and $\mhpm$ are much larger than $\mhl$
and they increase as $\mhl$ is increased.
Fig.~\ref{chi2min} illustrates this for the case $[\mha=90\gev,\tanb=2.3]$.
For all $\mhl$, $\dchisqmin$ is achieved for $\beta-\alpha\sim \pi/2$,
corresponding to maximal coupling of the lightest scalar $\hl$ to $ZZ$.
However, as to be expected from Fig.~\ref{m3mc}, if we look for all
solutions with $\dchisq<\dchisqmin+0.05$, many more choices
of closely correlated $\mhpm,\mhh$ pairings are possible with
much lower overall mass scale and these do not always have 
$\beta-\alpha\sim \pi/2$. Fig.~\ref{chi2min} also shows that if
we decrease the minimum value allowed for $\mhl$ to $\rts-\mz=410\gev$
(the value for which $Z\hl$ production is forbidden when $\rts=500\gev$),
then $\dchisqmin$ decreases from $\sim 0.86$ to $\sim 0.71$.
Since the $b\anti b\hl$ cross section is very small for 
$\mhl=410\gev$ and $\tanb=2.3$ and $\nu\anti\nu \hl$ is very slow to 
turn on, this would still be consistent with non-observation of the $\hl$.

In the case of $\h=\hl$, the analogue of the above discussion applies
except that we fix $\alpha=\beta$ (for zero $ZZ\h$ and $WW\h$ couplings), 
implying that the next lightest Higgs, which is always $\H=\hh$ for
small $\dchisq$, is SM-like. One achieves $\dchisqmin$
for $\mhh=\rts-10\gev$ and for small positive $\mhpm-\mha$
with $\mha$ and $\mhpm$ typically substantially larger than $\mhh$.
Again, if we force $\mhh$ to larger values for any given $[\mha,\tanb]$
choice, then $\dchisqmin$ increases as do the associated $\mhpm,\mha$ masses.

Finally, we note that sensitivity of $\dchisqmin$ to precise
inputs is small. For example, by switching
from fixed $m_b$ to running $m_b$  in the
Higgs loop graphs (a change that is part of the two-loop corrections), 
the minimum $\chi^2$ values achievable in the SM-like
limit and in the no-discovery scenarios both decrease 
by about 0.4 for the running mass
choice, leaving $\dchisqmin$ almost the same.  
Changes in input values of $\mt$, $\alpha_s$
and $\mw^{\rm LEP}$ also lead to compensating changes.
We believe $\dchisqmin$ to be stable against such
changes/choices to within 0.1. More sensitive are the pulls and the precise
heavy Higgs masses that yield $\chi^2_{\rm min}$. While small values for the
splitting between the heaviest neutral and the charged Higgs masses
are always preferred, the overall mass scale at $\chi^2_{\rm min}$
is quite sensitive to precise inputs.  

{\bf 4. Qualitative discussion.} 
It is useful to understand in an approximate analytic fashion
how it is that there need not be a light scalar
Higgs boson with substantial $ZZ$ coupling in order for
predictions for all the precision observables to be in good
agreement with data (see also \cite{maria}). Consider the quantity
$\Delta\rho$ which must be very small for a low $\chi^2$ fit~\cite{drhorefs}.
It is convenient to write 
\bea
\Delta\rho&=&{\alpha\over 16\pi s_W^2\mw^2}\Bigl(\cos^2(\beta-\alpha)
\bigl[f(\mha,\mhpm)+f(\mhpm,\mhl)-f(\mha,\mhl)\bigr]\nonumber\\
&+&
\sin^2(\beta-\alpha)\bigl[f(\mha,\mhpm)+f(\mhpm,\mhh)-
f(\mha,\mhh)\bigr]\Bigr)\nonumber\\
&+& \cos^2(\beta-\alpha)\Delta\rho_{SM}(\mhh)+\sin^2(\beta-\alpha)
\Delta\rho_{SM}(\mhl)\,,
\label{drhoform}
\eea
where
\beq
f(x,y)={x^2+y^2\over 2}-{x^2y^2\over x^2-y^2}\log {x^2\over y^2}
\label{fform}
\eeq
is symmetric in $x\leftrightarrow y$, vanishes for $x=y$ 
and is $>0$ for $x\neq y$, and
\beq
\Delta\rho_{SM}(m)=-{3\alpha\over 16\pi s_W^2\mw^2}
\left[f(m,\mw)-f(m,\mz)\right]-{\alpha\over 8\pi c_W^2}\,.
\label{drhosmform}
\eeq
The usual SM-like result, $\Delta\rho=\Delta\rho_{SM}(\mhl)$, 
is obtained by taking $\mha=\mhh=\mhpm$
and $\sin^2(\beta-\alpha)=1$ (for which $\hl$ carries all the $ZZ$
coupling). 

Consider the case of $\h=\ha$ and $\H=\hl$, for which $\mhl$ must
be large (to forbid $b\anti b \hl$ production) and $\mhpm-\mhh$
should be small and positive for a good $\chi^2$. For cases such
that $\dchisqmin$ is achieved with $\sin^2(\beta-\alpha)\sim 1$,
a simple expression can be given for $\Delta\rho$. 
Using Eqs.~(\ref{drhoform}) and (\ref{drhosmform}), 
and expanding appropriately, we find 
\beq
   \Delta \rho=\frac{\alpha}{16 \pi m_W^2 c_W^2}\left\{\frac{c_W^2}{s_W^2}
   \frac{m_{H^\pm}^2-m_{H^0}^2}{2}-3m_W^2\left[\log\frac{m_{h^0}^2}{m_W^2}
   +\frac{1}{6}+\frac{1}{s_W^2}\log\frac{m_W^2}{m_Z^2}\right]\right\}
\label{drhonew}
\eeq
From this expression, we see that the (negative) logarithmic increase
with increasing $\mhl^2$ (that is the same as present in the 1HDM SM)
can actually be canceled by a small (positive) difference between $\mhpm^2$
and $\mhh^2$, which is precisely the nature of the $\dchisqmin$
arrangements we find.  It is apparent from this result
that actual degeneracy, $\mhl\simeq\mhh\simeq\mhpm=M$, would
lead to bad logarithmic growth of $\Delta\rho$ with $M$
(as bad as in the SM). We also note
that small $\Delta\rho$ does not require $\cos^2(\beta-\alpha)\sim 0$;
cancellations similar to those illustrated above can still be arranged
in the more general case, although they are more complicated.  Finally,
for the case of a light $\h=\hl$, $\Delta\rho$ takes the form given above with
$\mhh^2\to\mha^2$ and $\mhl^2\to\mhh^2$,
and corresponding cancellations can be arranged by having a small positive
$\mhpm^2-\mha^2$.

Another observable of interest is the $S$ parameter. For light $\h=\ha$
($\hl$), small $\mhpm^2-\mhh^2$ ($\mhpm^2-\mha^2$) is always needed for
good $\chi^2$ fits.  In this limit, independent of the size of
$\mhh^2-\mhl^2$ ($\mha^2-\mhh^2$) (and independent of $\sin^2(\beta-\alpha)$
in the $\h=\ha$ case to the extent that $\mha\sim\mw$) we find
\beq
S(0)\sim {1\over 12\pi}\left(-{5\over 3}+\log{\mH^2\over \mw^2}\right)\,,
\label{slimit}
\eeq
where $\H=\hl$ ($\H=\hh$), respectively. The fact that $S$ grows
logarithmically implies that the $\chi^2$ of the precision electroweak
fit will slowly worsen with increasing mass $\mH$ of the lightest kinematically
inaccessible Higgs boson, as noted, for example, in discussing
the $[\mha=90\gev,\tanb=2.3]$ case.
The non-decoupling effects in $S$ (and other observables) of large $\mH$
imply that as $\mH$ is forced to increase with $\rts$ (as required
to avoid $\H$ discovery in the $Z\H$ channel if not the $b\anti b\H$
channel) the $\dchisqmin$ achievable will slowly increase relative to
the SM-like limit of a light $\hl$ with full $WW,ZZ$ coupling strength.

{\bf 5. The Higgs potential.} 
It is interesting to ask if 
the parameter configurations that minimize $\dchisq$
could find explanation in symmetries at the Lagrangian level.
After imposing the usual $Z_2$ symmetry
to avoid hard flavor changing terms in the
Lagrangian, the 2HDM potential can be written
in terms of the two SU(2) Higgs doublets
$\Phi_1=(\phi_1^+,\phi_1^0)$ and $\Phi_2=(\phi_2^+,\phi_2^0)$ in the
form (see, e.g., \cite{ggk}):
\begin{eqnarray}
 V(\po,\pht)&=&-\mu_1^2\popo-\mu_2^2\phtpt
-(\mu_{12}^2\popt+\hc)   
+{1\over 2}\lambda_1(\popo)^2+{1\over 2}\lambda_2(\phtpt)^2
\nonumber \\
&+&\lambda_3(\popo)(\phtpt) +\lambda_4|\popt|^2+\frac{1}{2}
\left[\lambda_5(\popt)^2+\hc\right]\,,
\label{potform} 
\end{eqnarray}
where $\mu_{12}^2$ and $\lambda_5$ 
should be chosen real for a CP-conserving Higgs potential.
The resulting Higgs masses or mass matrices are then
\bea
\mha^2&=&{\mu_{12}^2\over \sb\cb}-v^2\lam_5\,, \quad
\mhpm^2=\mha^2+{1\over 2}v^2(\lam_5-\lam_4)\nonumber \\
{\cal M}^2&=&\mha^2\left(\begin{array}{cc} \sb^2 & -\sb\cb \\
-\sb\cb & \cb^2 \\ \end{array}\right)
+v^2\left(\begin{array}{cc} \lam_1\cb^2+\lam_5\sb^2 & (\lam_3+\lam_4)\sb\cb \\
(\lam_3+\lam_4)\sb\cb & \lam_2\sb^2+\lam_5\cb^2 \\ \end{array}\right)\,.
\label{masssq}
\eea
So long as $\mha^2>0$, the CP-conserving minimum is either
the only minimum ($\lam_5>0$) or the preferred minimum ($\lam_5<0$).
Let us  look first for relations among the $\lam$'s that will
lead to the $\mhpm\gsim\mhh>\mhl\gg\mha$ and $\beta-\alpha\sim \pi/2$
configuration. Consider the case where $\lam_1=\lam_2$. 
To obtain $\beta-\alpha=\pi/2$
we must, first of all, have $\tan2\alpha=\tan2\beta$, which requires
$\lam_1-\lam_5=\lam_3+\lam_4$. Then, $\mhl^2=\lam_1 v^2$ and
$\mhh^2=\mha^2+\lam_5 v^2$. In the limit of small $\mha$,
we then require $\lam_1<\lam_5$ for $\mhl<\mhh$. Further,
for $\mhpm\sim \mhh$ when
$\mha$ is small, we need $\lam_4=-\lam_5$.
Then, $\lam_1-\lam_5=\lam_3+\lam_4$ 
implies $\lam_1=\lam_3$. 
Meanwhile, $\mha^2$ will be small if $\mu_{12}^2\sim \lam_5 \sb\cb v^2$.
Thus, our potential should have
\beq
\lam_1\sim\lam_2\sim\lam_3>0\,,\quad \lam_5=-\lam_4>0\,,\quad \lam_5>\lam_1\,. 
\label{lams}
\eeq  
This configuration is achieved for a rather amusing
special form of the quartic terms in the potential:
\beq
V_{\rm quartic}(\po,\pht)={1\over 2}\lam_1\left|\popo+\phtpt\right|^2-
{1\over 2}\lam_5
\left|\popt-\phtpo\right|^2\,,
\label{vquartic}
\eeq
i.e. a weighted sum of the (absolute) squares
of the natural symmetric and antisymmetric combinations of
the two Higgs doublet fields. This form of the potential guarantees
absence of quadratic growth of $\Delta\rho$ with
the masses of the heavier Higgs bosons, i.e. it incorporates a hidden
custodial SU(2) symmetry. 
We need only break these various relations by relatively small amounts
in order to achieve  the small $\mhpm-\mhh$ mass difference
required for good $\chi^2$ for the precision electroweak fit.
Of course, when $\lam_5>\lam_1>0$, $V$ is globally unstable with 
respect to large (CP-violating) field directions, $\phi_1^0= ix_1$
and $\phi_2^0=x_2$ ($x_1$ and $x_2$ real) with $x^2\equiv x_1^2+x_2^2\to\infty$
and $4\lam_5x_1^2x_2^2/x^4>\lam_1$.  However, it has been argued
\cite{vacuum} that such globally bad directions are not likely to have been
entered during the evolution of the universe following
the big bang so long as
the local ($\mha^2>0$) stable minimum is nearer to the origin in
field value space than the globally unstable regions. Thus, for $\mha>50\gev$
the global instability of the potential 
form that minimizes our precision electroweak $\chi^2$ should probably not
be regarded as a problem.

To achieve $\alpha\sim 0$ and $\mhpm\sim\mhh\sim\mhl\gg\mha$,
one should choose $\lam_1=\lam_2=\lam_3=\lam_5=-\lam_4$, i.e. the $\lam_1=\lam_5$ limit of the above form (which has only a globally flat limit).
Small deviations from this form would then allow for the $\mhpm>\mhh\gsim\mhl$
configuration required for minimum $\chi^2$ when $\alpha\sim 0$. 
Finally, if $\alpha=\beta$ we have $\mhl^2=\mha^2+v^2\lam_5$
and $\mhh^2=v^2\lam_1$. Then, $\mhpm\sim\mha\sim\mhh\gg\mhl$ provided
$\lam_1=\lam_2=\lam_3=-\lam_5=-\lam_4=\mha^2/v^2$,
yielding a potential form with no global problems.

{\bf 6. Higher energy LC and LHC.}
Having delineated the dangerous choices of parameters for which 
failure to detect a Higgs boson at a $\rts\leq 800\gev$ $\epem$
collider can be rather consistent with precision electroweak constraints,
we wish to address the question of whether increased LC energy or LHC
running 
{\it would} allow Higgs discovery.  First, by comparing the $\rts=500\gev$
and $\rts=800\gev$ no-discovery wedges, we see that although
the $\tanb$ extent of the wedge narrows considerably
with increasing $\rts$, the smallest no-discovery
value of $\mh$ increases rather slowly; thus, one cannot absolutely
rely on $\h$ detection at higher LC energy in these scenarios.
Also, the absence of $ZZ$ coupling and the moderate value
of $\tanb$ implies that the $\h$ will not be detectable at the LHC.
Consider next the $\H$. Since $\chi^2_{\rm min}$ is always
achieved for $\mH$ at the 
$\H b\anti b$ threshold 
and for masses of the other Higgs bosons often
much larger than $\mH$, the discovery possibilities for the 
$\H=\hl$ ($\hh$) deserve particular attention in the 
$\h=\ha$ ($\hl$) cases.

Consider first the cases where $\h=\ha(\hl)$ and $\dchisqmin$ is
achieved when the $\H=\hl(\hh)$ is SM-like. For the $\dchisqmin$ values of
$\mH$ and for a substantial range above, the LHC would detect the $\H$
in the gold plated $ZZ\to 4\ell$ channel.  If $\rts$ of the $\epem$
collider is pushed beyond 1 TeV, and the $\H$ is not seen in $Z\H$ or
$\nu\anti\nu\H$, the minimum $\mH$ would move into the $\gsim 1\tev$
range for which one would expect to see strong $WW$ scattering
behavior at both the LHC and the LC. Our results indicate that
precision electroweak fits do not necessarily have particularly bad
$\chi^2$ for such large $\mH$ --- typically, $\dchisqmin$ only
increases by $<1-2$ compared to values obtained for $\mH\sim
800\gev$.\footnote{This  conclusion is not entirely
iron-clad for the following reason. The heavy Higgs masses are given
by $\sqrt{\lam} v$, where $\lam$ is some combination of the $\lam_i$. 
To obtain masses of order $1\tev$ (as typical for good $\chi^2$
in the $\rts=800\gev$ wedge), we need $\sqrt\lam\sim 4$ or 
$\lam^2/(4\pi)\sim 1$. Thus, 
our perturbative precision electroweak computations
undoubtedly begin to break down for $\mH$ values much above
$800-900\gev$.}
As a result, only an LC with $\rts$ sufficiently large
to probe a strongly interacting $WW$ sector could be certain of 
seeing a signal for the $\H$. Note that although the $b\anti b\H$
production channel opens up as the $\rts$ of the LC is increased,
$\sigma(b\anti b\H)$ for a SM-like $\H$ is very small at high mass and 
$b\anti b\H$ production would not be detectable.

Continuing to focus on the $\h=\ha$, $\H=\hl$ SM-like with 
$\sin^2(\beta-\alpha)\sim 1$ 
situation, the two heaviest Higgs
bosons $\hh$ and $\hpm$ have fairly large masses
(e.g. $600-800\gev$ at a $\rts=500\gev$ LC) for the best $\chi^2$
values.  Although the $Z\to \ha\hh$
coupling would have full strength, the masses are such that $\ha\hh$
production would become kinematically allowed only with a substantial increase
in $\rts$.  Because of the small cross sections for Yukawa processes
at moderate $\tanb$, much larger $\rts$ would be needed for $b\anti b\hh$
and $b\anti t\hp+\anti b t \hm$ production. And, much larger $\rts$
would also be required for $\hp\hm$ and $t\anti t\hh$ production.
For $\rts=800\gev$, the $\dchisqmin$ cases with $\sin^2(\beta-\alpha)\sim 1$
have $\mhh$ and $\mhpm$ above 1 TeV. A $\rts>2\tev$ LC would be needed
to discover these Higgs in pair production.  Because of the moderate
value of $\tanb$, this would also be true for Yukawa processes.  For
moderate $\tanb$ and masses as large as discussed above, $\hh$ and $\hpm$
detection at the LHC would not be possible due to the smallness of the $ZZ\hh$
and $WW\hh$ couplings and the very modest size of $b\anti b\hh$ production.
Thus, in these $\dchisqmin$ cases, the
first focus should be on LHC observation of the $\hl$ as a resonance
or in strong $WW$ scattering.

Next consider the $\h=\ha$ cases for which $\dchisqmin$ is achieved
for small $\sin^2(\beta-\alpha)$, as typified by the moderate $\tanb$, 
$\alpha\sim 0$ cases.  The $\H=\hl$ will be hard to detect in the
SM-like discovery modes.  However, $\ha\hl$ production would
have more or less full strength coupling and observation would be
possible when kinematically allowed. Since our searches were performed
subject to the requirement $\rts<\mhl+10\gev$, this would typically require
very substantially larger $\rts$ than the assumed value.
However, in these cases the $\hh$ has SM-like $ZZ,WW$ coupling and
$\mhh$ is usually not much larger than $\mhl$ (which is always $\rts
-10\gev$). Consequently, $\hh$ detection in the gold-plated modes
at the LHC or at a $\rts\gsim 1\tev$ LC would be possible.

Finally, for the $\h=\hl,\H=\hh$ (with $\hh$ SM-like) case, the two
heaviest Higgs bosons are the $\ha$ and $\hpm$. As $\rts$ at the LC is 
increased, $\hl\ha$ production would become kinematically allowed
(and be full strength), followed by
$\hp\hm$ pair production. For the moderate $\tanb$ values in question, the
Yukawa processes would not be useful (either at the LC or the LHC).

{\bf 7. Conclusions.} Although it is likely that a light SM-like
Higgs boson exists and will be found at the LHC and LC if
not in the near future at 
LEP2 or the Tevatron, it is advisable to consider scenarios in
which the Higgs boson(s) with substantial $WW/ZZ$ coupling are heavy. 
We have shown that even the very excellent level of precision achieved
for electroweak observables is not adequate to clearly 
prefer a Higgs sector (either 1HDM or 2HDM)
with a light SM-like Higgs over a type-II 2HDM 
model with parameters chosen so that no Higgs boson will be discovered
at a $\rts=500-800\gev$ machine in $\epem\to Z\h$, 
$\nu\anti\nu\h$, $b\anti b\h$
or $t\anti t\h$. In particular,
we have considered the case of the CP-conserving
2HDM model with a light pseudoscalar $\h=\ha$
or scalar $\h=\hl$ with zero $ZZ/WW$ coupling and
all other Higgs bosons too heavy to be produced.
For moderate $\tanb$, there is a range of $\mh$ for which the $\h$
will typically not be detectable even
with very high integrated luminosity and yet the masses
of the other Higgs bosons can be chosen to be $\gsim \rts$
{\it and} so that the precision electroweak
fit is within $\Delta\chi^2=1-2$ of the best SM fit.
Although some tuning of parameters is required for the smallest
$\Delta\chi^2$, the parameter relations required are small
perturbations relative to
an interestingly simple form of the Higgs potential.
Fortunately, Giga-$Z$ electroweak precision measurements
will allow much stronger discrimination
between the light SM-like Higgs and the `no-$\epem$-discovery' 2HDM scenarios,
while $\gam\gam$ collisions will allow $\h$ discovery for $\mh<0.8\rts$
for some portion of the low to moderate $\tanb$ regions
of the no-discovery wedges, depending upon the effective $\gam\gam$
integrated luminosity.

\vspace{.2cm}
\centerline{\bf Acknowledgments}
\vspace{.2cm} 
This work was supported in part by the Committee for
Scientific Research (Poland) under grants No. 2~P03B~014~14, 2 P03B 052 16,
and 2~P03B~060~18, by the U.S.
Department of Energy under grant No. DE-FG03-91ER40674
and by the U.C. Davis Institute for High Energy Physics.
JG thanks the Aspen Center for Physics, JK
thanks  the DESY Theory group and BG thanks the SLAC Theory group
for hospitality during completion of this
work. JK acknowledges many interesting discussions with P. Zerwas.


\begin{thebibliography}{99}
\bibitem{latestdata} For all quantities except $\mw^{\rm LEP}$ we employ
the results contained in LEPEWWG CERN-EXP-00-16.
\end{thebibliography}
\end{document}